# Intensity-based dynamic speckle method for analysis of variable-rate dynamic events

Elena Stoykova, Lian Nedelchev, Blaga Blagoeva, Branimir Ivanov, Mikhail Levchenko, Nataliya Berberova-Buhova, Dimana Nazarova
Institute of Optical Materials and Technologies, Bulgarian Academy of Sciences,
Acad. Georgi Bonchev Str., Bl.109, 1113 Sofia, Bulgaria;

## ABSTRACT

We study efficiency of intensity-based dynamic speckle method for characterization of dynamic events which occur at variable rate in time within the temporal averaging interval. We checked ability of the method to describe the speed evolution by i) numerical simulation at variable speed, ii) processing of speckle patterns obtained from phase distributions fed to a SLM at controllable change of the temporal correlation radius of speckle intensity fluctuations and iii) conducting experiments with a polymer solution drying by using a hot-stage. The numerical and SLM simulation experiments allowed for modification of the used estimates in order to obtain relevant information.

## 1. INTRODUCTION

Intensity-based dynamic speckle method (IDSM) is a reliable tool for evaluating speed of processes ongoing in objects of biological or industrial origin[1-4]. Pointwise implementation of the method by using a 2D optical sensor gives spatio-temporal activity distribution extracted from a set of correlated in time speckle patterns formed on an object illuminated with coherent light[5,6]. Speckle nature of the raw data demands averaging to be applied to build a reliable estimate of the speed. In the ideal case, when spatial or temporal averaging is applied, speed is assumed constant within a certain area or a temporal interval respectively. Tracking temporal variation of speed at each object point entails a set of activity maps to be built using sequential temporal intervals for averaging that may partially overlap or have abutting borders. An activity map depicts regions of slower or faster intensity fluctuations in the speckle patterns on the object surface and, after calibration, can characterize quantitatively the spatial speed distribution.

      This study aims at developing an IDSM modification for efficient characterization of dynamic events which occur at variable rate in time during capture of the speckle images. Such a situation is observed when the acquisition rate of speckle images is comparable to the speed change. We checked ability of the IDSM to describe the speed evolution by i) numerical simulation at variable speed, ii) processing of speckle patterns obtained from phase distributions fed to a SLM at controllable change of the temporal correlation radius of speckle intensity fluctuations and iii) conducting experiments with a polymer solution drying by using a hot-stage. The numerical and SLM simulation experiments allow for modification of the estimates in order to obtain relevant information. The results from the drying experiment at higher temperature illustrate capability of the modified ISDM to track fast-developing processes.

## 2. IDSM SIMULATION AT VARIABLE RATE OF SPECKLE DYNAMICS

### 2.1 Numerical simulation scheme

We consider the IDSM which is realized with a 2D optical sensor, focused on a 3D object surface. The sensor captures correlated in time speckle patterns of size $N_x \times N_y$ pixels at a pixel interval, $\Delta$. The time interval between the consecutive images is $\Delta t$. For each pixel, $(k,l) \equiv (k\Delta, l\Delta), k=1..N_x, l=1..N_y$, a temporal sequence of intensity values, $I_{kl,n} = I(k\Delta, l\Delta, n\Delta t), n=1..N$, is composed from $N$ images. For 8-bit encoded images, the intensity values are integers from 0 to 255. The recorded intensities are input data for pointwise correlation-based algorithms, which estimate temporal correlation between images separated by the time lag, $\tau = m\Delta t$, where $m$ is an integer. The spatial distribution of the estimate is called an activity map. Usually, the estimate,

$$S_1(k,l,m) = \frac{1}{(N-m)} \sum_{n=1}^{N-m} |I_{kl,n} - I_{kl,n+m}| \quad (1)$$

gives reliable maps for symmetrical/asymmetrical speckle intensity distributions. This pointwise estimate, as many others, is non-linearly related to the spatial distribution of the temporal correlation radius, $\tau_c(k,l)$, of intensity fluctuations on the object surface. The drop in $\tau_c(k,l)$ means rise of activity and increase of $S_1$ respectively. The estimate (1) is obtained by averaging over time. Its spatial resolution is given by the pixel pitch, but this advantage is partially compromised by the fact that the estimate strongly fluctuates from point to point thus worsening the map contrast. The condition for a good contrast map is to provide as narrow as possible probability density function or histogram of the estimate at a given $\tau_c$. The spread of this histogram decreases with averaging interval $T = N\Delta t$ which can exceed or be less than $\tau_c(k,l)$ at different object points. The estimate (1) relies on the assumption that $\tau_c(k,l)$ remains unchanged during the capture of the used *N* patterns.

In this study, we consider the case when the temporal correlation radius depends also on time, so $\tau_c = \tau_c(x,y,t)$ is fulfilled. More specifically, we assume that the acquisition rate is comparable to the rate of change of $\tau_c$, so any new captured image at $t = n\Delta t, n = 1,2...N$ corresponds to a new distribution $\tau_c(x,y,t=n\Delta t)$ in space. For the simulation, we chose the following model:

$$\tau_c(x=k\Delta, y=l\Delta, t=n\Delta t) \equiv \tau_c(k,l,n) = 260\Delta t - (n-1)\Delta t \times \exp\left\{-\frac{(k\Delta-128\Delta)^2 + (l\Delta-128\Delta)^2}{(112\Delta)^2}\right\} \quad (2)$$

with $N_x \times N_y = 256 \times 256$ and $n = 1,2...256$. For this model, activity increases in time being maximal at the center of the image. The object is illuminated with an expanded laser beam with uniform intensity distribution at wavelength, $\lambda$. A process ongoing in the object causes random movement of the scattering centers on the rough object surface. For the purpose of simulation, it is enough to assume that these centers move randomly in the two directions normal to the object surface. The complex amplitudes of light scattered from the centers have mutually independent amplitudes and phases for a given center and between the centers. No temporal change in reflectivity is observed across the object during acquisition of the raw data. Thus the phase change due to the height change of a scattering center is normally distributed at each point[7]. The phase change, $\Delta\varphi_m^{kl,n}$, at point $(k\Delta, l\Delta)$ between the moments $n\Delta t$ and $(n+m)\Delta t$ separated by a time lag $\tau = m\Delta t, m = 1,2...N_\tau < N$ leads to intensity fluctuations in the optical sensor with a normalized correlation function[7,8] $\rho_{kl,n}(\tau = m\Delta t) = \exp(-\sigma^2\{\Delta\varphi_m^{kl,n}\})$, where $\sigma^2\{\Delta\varphi_m^{kl,n}\}$ is the variance of the phase change. We used the model $\rho_{kl,n}(\tau) = \exp[-\tau/\tau_c(k,l,n)]$. This model describes effectively many processes as e.g. a drying process. We assume that for the images (*n*-1) and *n* the following formula is acceptable for the standard deviation of the phase change, $\sigma\{\Delta\varphi_{m=1}^{kl,n}\} = \sqrt{\Delta t/\tau_c(k,l,n)}$, where $n\Delta t$ is the time instant of recording the *n*-th image. As a first step, delta-correlated in space random phases $\varphi(k\delta, l\delta, n\Delta t), k = 1..2N_x, l = 1..2N_y, n = 1..N$ were generated on the object surface at a spatial step $\delta = \Delta/2$ starting from a 2D array of phase values uniformly distributed from 0 to $2\pi$. The phase distribution at instant $n\Delta t$ for $n \geq 2$ is found from $\varphi(k\delta, l\delta, n\Delta t) = \varphi[k\delta, l\delta, (n-1)\Delta t] + \chi_{kl,n}\sqrt{\Delta t/\tau_c(k,l,n)}$ where $\chi_{kl,n}$ is a random number with standard normal distribution with zero mean and variance equal to 1, $k = 1..2N_x, l = 1..2N_y, n = 1..N$. This number is separately generated for each combination of indices, $k,l,n$. The complex amplitude on the object surface was generated from $\varphi(k\delta, l\delta, n\Delta t)$ for intensity distribution $I_0(k\delta, l\delta)$ of the laser beam on the object surface as $U_S = \sqrt{I_0(k\delta, l\delta)}\exp\{-j[\varphi(k\delta, l\delta, n\Delta t)]\}$, where *j* is the imaginery unit. At this stage of simulation, spatial intensity distribution $I_0(k\delta, l\delta)$ of the laser beam was given by real numbers. Then the complex amplitude of the light field on the sensor aperture was generated as $U_{cam} = FT^{-1}\{H \cdot FT\{U_S\}\}$ where $FT\{\}$ denotes Fourier transform and *H* is a *circ* function in the Fourier domain with a cut-off frequency equal to $N_x\delta D/(2\lambda f)$, where *D* and *f* are the diameter and the focal distance of the camera objective and we assumed[9] that $N_x = N_y$. Integration of speckle by the camera pixels was done by summation of values $|U_{cam}|^2$ in a window of size 2×2 pixels. It was assumed that the exposure interval was much shorter than $\Delta t$, and no averaging of the speckle within this interval was simulated.

## 2.2 Analysis of variable-rate dynamic events

We generated 256 speckle patterns following model (2) for the temporal correlation radius. Figure 1(a) depicts spatial distribution of $\tau_c(x, y, t = 240\Delta t)$. Note that the temporal correlation function of intensity fluctuations reflects non-linearly this distribution. The wavelength was 532 nm. We processed the generated images by applying estimate (1) for $N = 64$ and choosing different initial moments, $t_i$. Three of the obtained activity maps are presented in Fig.1(b). The maps are plotted using the same scale for better comparison, and the numbers along the horizontal and vertical axes correspond to pixels. Although the temporal interval $T = 64\Delta t$ is rather short as an averaging interval, it is quite long with respect to changes in $\tau_c(x, y, t)$, so the estimate fails in adequate description of activity at different moments. Only for the sequence of images starting at $t_i = 190\Delta t$, some correspondence to the activity distribution is observed.

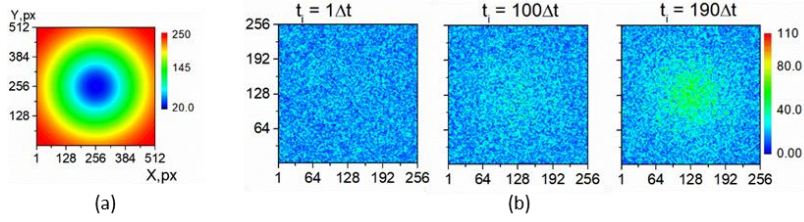

Figure 1. Spatial distribution of the temporal correlation radius for the model given by Eq.(2) at the instant $240\Delta t$ (a); activity maps of estimate (1) corresponding to different initial instants in the processed sequences of intensity; time lag $\tau = 15\Delta t$, image size 256×256 pixels, uniform illumination at 532 nm (b).

To improve evaluation of activity, when the process is fast-developing, averaging must be performed in the spatial domain. One may divide each image in segments of equal size and calculate the estimate for a certain time lag using only two images and averaging within a segment. The estimate built by following such a procedure for square segments is given by:

$$S_2(K, L, n, m) = \frac{1}{ss^2} \sum_{k=(K-0.5)ss}^{(K+0.5)ss} \sum_{l=(L-0.5)ss}^{(L+0.5)ss} |I_{kl,n} - I_{kl,n+m}| \quad (3)$$

where $(K,L)$ are the indices of the segment of size $ss \times ss$ pixels with $K = 1, 2 ... N_x / ss; L = 1, 2 ... N_y / ss$, and $n$ and $m$ indicate the time instant and the time lag respectively. The segments have abutting boundaries. This estimate enables better tracking of activity and is suitable for fast calculation of the activity map. However, we proposed another modification of the estimate in which the averaging segment is sliding within the image as follows:

$$S_3(k', l', n, m) = \frac{1}{ss^2} \sum_{i=k'+1}^{k'+ss} \sum_{j=l'+1}^{l'+ss} |I_{ij,n} - I_{ij,n+m}|; k' = k - ss/2, l' = l - ss/2 \quad (4)$$

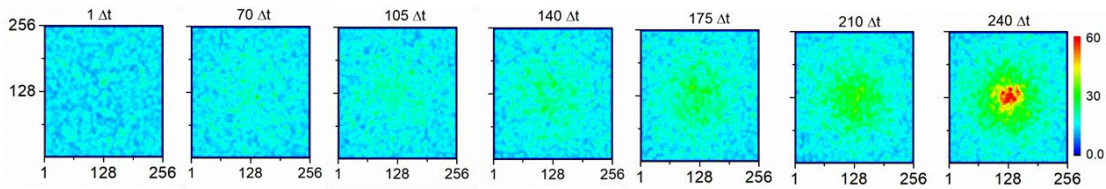

Figure 2. Activity maps for estimate (4) at different instants for the model given by Eq.(2) and a time lag equal to $\tau = 15\Delta t$; image size 256×256 pixels, segments size 8×8 pixels, uniform illumination at 532 nm.

Activity maps for estimate (4) at different instants and time lag $\tau = 15\Delta t$ are shown in Fig.2. Rise in activity in time is clearly seen. Estimates $S_2$ and $S_3$ depend on time and the time lag. Actually, the captured set of 256 images allows for visualization of activity at different instants and time lags in order to characterize fully the rate and location of changes within the object. The small time lags enable more accurate tracking of the speed change. However, they lead to maps with a low contrast, so it is difficult to retrieve the relevant information due to strong fluctuations of the estimates from point to point. Not to lose information at small and big time lags, one can built the contour maps for the estimates as a function of the time lag and time at different points of the object surface. Such maps for the model given by Eq.(2) and estimate $S_3$ are presented in Fig.3.

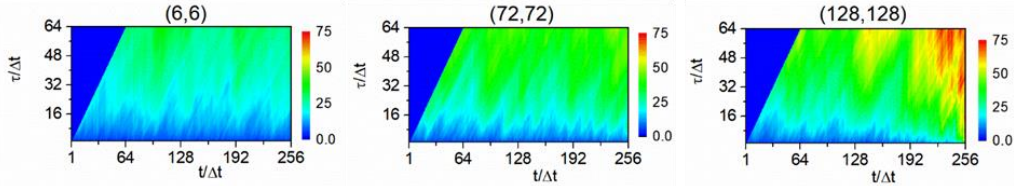

Figure 3. Contour maps for estimate (4) as a function of time and the time lag for the model given by Eq.(2) at points $(6\Delta,6\Delta)$, $(72\Delta,72\Delta)$ and $(128\Delta,128\Delta)$; image size 256×256 pixels, segments size 8×8 pixels, illumination at 532 nm.

**2.3 SLM simulation**

As a second simulation, we generated correlated phase distributions and fed them to a phase-only SLM with a phase span of 2π. A sequence of speckle patterns were imaged on a screen under illumination at a wavelength 632.8 nm and captured by the optical sensor. The patterns were generated for a rectangular area with changing in time temporal correlation radius. This area was surrounded by a background with large and constant temporal correlation radius. The high activity area is indicated by the white rectangle in Fig.4 (a) which depicts an exemplary speckle patter of size 600 by 600 pixels chosen for processing.

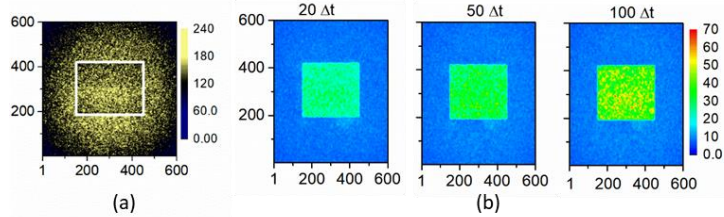

Figure 4. Exemplary speckle pattern produced by a phase-only SLM (a); activity maps at different instants for a time lag equal to $\tau = 15\Delta t$; image size 600×600 pixels, segments size 8×8 pixels, wavelength 632.8 nm (b).

## 3. EXPERIMENTAL CHECK

To check applicability of the developed approach, we recorded a sequence of speckle images for a drying experiment of a polymer solution. We used the azopolymer poly [1- [4- (3-carboxy-4-hydroxyphenylazo) benzenesulfonamido] -1,2-ethanediyl, sodium salt] or shortly PAZO. The polymer is available from Sigma Aldrich[10] and its chemical structure is shown in Fig.5. PAZO is easily soluble in water and methanol. The methanol was chosen to create a fast-developing process. The methanol droplet was completely dry for less than 2 min at 30°C. A solutions was prepared for the drying experiment from 10 mg PAZO dissolved in 400 μL methanol. This concentration is suitable for the preparation of thin films from this azopolymer[10]. To monitor the drying of the solution, a 10 μl droplet was spread on a microscope glass slide. The glass was placed on a hot stage THMS 600 (Linkam Scientific). The stage allows for keeping accurately the sample temperature at a pre-set value. Thermal equilibrium was achieved by leaving the glass substrate for 5 minutes on the stage at the desired temperature before spreading the droplet. The temperatures applied were 30, 40 and 50°C. The exposure time was 20 microseconds and time Δt between the consecutive images was 250 milliseconds. The room temperature maintained and monitored by an air-conditioning system was 25°C throughout the experiment.

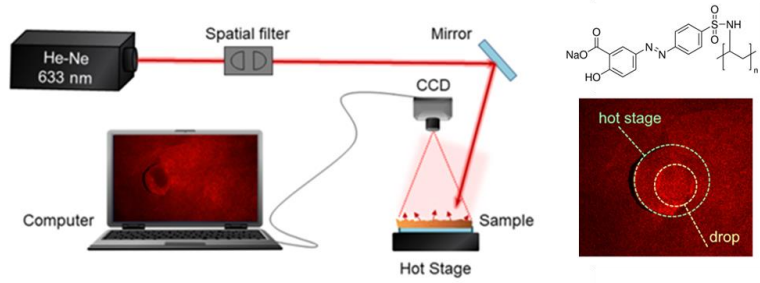

Figure 5. Experimental set-up with a hot-stage for IDSM implementation for monitoring of a drying process of the drop of azopolymer PAZO solution in methanol.

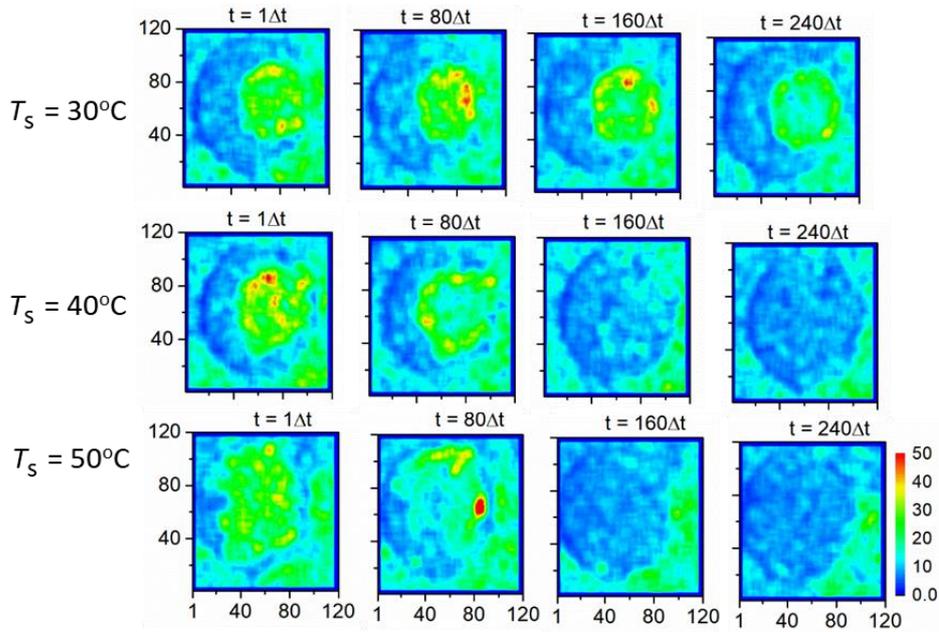

Figure 6. Activity maps for estimate (4) at different instants and temperatures for the polymer drop drying experiment and a time lag equal to $\tau = 10\Delta t$; image size 120×120 pixels, segments size 8×8 pixels, wavelength 632.8 nm.

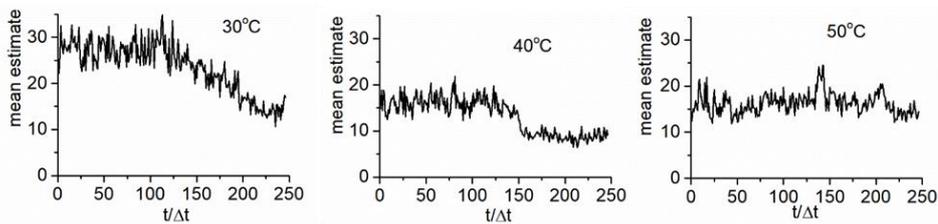

Figure 7. Decrease of the mean estimate (4) with time for a region of 11×11 segments corresponding to the azopolymer solution at different temperatures and a time lag equal to $\tau = 10\Delta t$; image size 120×120 pixels, segments size 8×8 pixels, wavelength 632.8 nm.

The experimental set-up is shown in Fig.5. A color CMOS camera with 780×582 pixels and a pixel pitch $\Delta$ = 8.1 μm was focused on the object with its optical axis normal to the object surface. A He-Ne laser emitting at wavelength 632.8 nm was chosen as a light source. The entire optical set-up was placed on a vibration-isolated table. As speckle is extremely sensitive to micro-variations of topography or refractive index, drying of the droplet inflicts changes in the speckle pattern formed on the surface under laser illumination. The results of processing the recorded sequences of speckle images are shown in Fig.6 and Fig.7. We did not present the activity maps for the pointwise processing by using estimate (1) because they are not informative. The activity maps in Fig.6 show decrease of activity in time: activity is the highest at the beginning of drying and goes down to almost zero activity. The decrease of activity is slow only for the temperature of 30°C. The maps are plotted with the same color scale for better comparison. To track decrease of activity in time, we found the mean of the estimate $S_3$ within a region of 11×11 segments. Each value of $S_3$ is calculated by averaging in a segment of size 8×8 pixels.

## 4. CONCLUSIONS

In summary, we developed a modified intensity-based algorithm for tracking dynamics of fast-developing processes in objects by using a sequence of correlated in time speckle patterns on their surface as input data. The algorithm output is a set of activity maps representing evolution of of spatial distribution of degree of correlation between two images separated by a given time lag. A given activity map is calculated by partitioning two images into overlapping segments with a small number of pixels. The estimate at a given point of the map is obtained by averaging within a segment. The neighboring value of the estimate is obtained from the next segment shifted horizontally or vertically at one pixel with respect to the first one. For fast processing, segments with abutting borders can be used. The dynamics can be characterized by processing at more than one time lag. The smaller the time lag, the better the temporal resolution of the algorithm. A small time lag, however, may lead to a low contrast map and poor visualization of activity due to the strongly fluctuating in space values of the estinate. Usage of a small segment size intensifies these fluctuations. The proposed algorithm was applied to simulated and experimental data with satisfactory results. For the experiment, the algorithm made possible visualization of the drying process of a drop of a polymer dissolved in a methanol. The obtained set of activity maps allowes for tracking the drying process with good temporal resolution. For comparison, if only averaging in time is applied, the result corresponds to constant lack of activity of a fully dried drop.

Acknowledgement

## REFERENCES

[1] S. E. Murialdo, G. H. Sendra, L. I. Passoni, R. Arizaga, J. F. Gonzalez, H. Rabal, and M. Trivi, "Analysis of bacterial chemotactic response using dynamic laser speckle," J. Biomed. Opt. 14(6), 064015 (2009).
[2] B. Mandracchia, J. Palpacuer, F. Nazzaro, V. Bianco, R. Rega, and P. Ferraro, "Biospeckle decorrelation quantifies the performance of alginate-encapsulated probiotic bacteria", IEEE Journal of Selected Topics in Quantum Electronics 25(1), 1-6 (2018).
[3] Chen, L., Cikalova, U., & Bendjus, B. (2018, September). Laser speckle photometry: an advanced method for defect detection in ceramics. In Speckle 2018: VII International Conference on Speckle Metrology (Vol. 10834, pp. 302-308). SPIE.
[4] A. van Welzen, J., Yuan, F. G., & Fong, R. Y. Hidden damage visualization using laser speckle photometry. NDT & E International, 131, 102700 (2022).
[5] A.V. Saúde, F. S. de Menezes, P. L. Freitas, G. F. Rabelo, and R. A. Braga, Jr., "Alternative measures for biospeckle image analysis," J. Opt. Soc. Am. A 29(8), 1648–1658 (2012).
[6] E. Stoykova, B. Ivanov, and T. Nikova, "Correlation-based pointwise processing of dynamic speckle patterns," Opt. Lett. 39(1), 115-118 (2014).
[7] T. Fricke-Begemann, G.Gülker, K. D. Hinsch, and K. Wolff, "Corrosion monitoring with speckle correlation," Appl. Opt. 38, 5948-5955 (1999).
[8] A. Federico, G. H. Kaufmann, G. E. Galizzi, H. Rabal, M. Trivi, and R. Arizaga, "Simulation of dynamic speckle sequences and its application to the analysis of transient processes," Opt. Commun. 260(2), 493–499 (2006).
[9] E. Equis and P. Jacquot, "Simulation of speckle complex amplitude: advocating the linear model," Proc. SPIE 6341, 634138 (2006)

[10] Nedelchev, L., Mateev, G., Otsetova, A., Nazarova, D., Stoykova, E. "Optimization of deposition of thin photoanisotropic films for holographic data storage," Int. J. Inf. Theor. Appl. 25, 245–254 (2018).